\documentclass[%
 reprint,
superscriptaddress,
 amsmath,amssymb,
prb,
]{revtex4-1}

\usepackage{amsmath}
\usepackage[utf8]{inputenc}
\usepackage{siunitx} 
\usepackage{xcolor}
\usepackage{natbib}
\usepackage{graphicx}
\usepackage{braket}
\usepackage{comment}
\usepackage{bm}
\usepackage{epstopdf}

\usepackage{graphicx}
\usepackage{dcolumn}
\usepackage{bm}
\usepackage{comment}

\usepackage{braket}
\usepackage{color}
\usepackage{array}
\newcolumntype{?}{!{\vrule width 1pt}}

\begin{document}
%
%



\title{GaAs quantum dots under quasi-uniaxial stress: experiment and theory}

\date{\today}

\author{Xueyong Yuan}
\email{xueyongyuan@seu.edu.cn}
\affiliation{School of Physics, Southeast University, 211189 Nanjing, China}
\affiliation{Institute of Semiconductor and Solid State Physics, Johannes Kepler University Linz, Altenbergerstra{\ss}e 69, A-4040 Linz, Austria}

\author{Saimon F. Covre da Silva}
\affiliation{Institute of Semiconductor and Solid State Physics, Johannes Kepler University Linz, Altenbergerstra{\ss}e 69, A-4040 Linz, Austria}

\author{Diana Csontosov\'{a}}
\affiliation{Department of Condensed Matter Physics, Faculty of Science, Masaryk University, Kotl\'a\v{r}sk\'a~267/2, 61137~Brno, Czech~Republic}

\author{Huiying Huang}
\affiliation{Institute of Semiconductor and Solid State Physics, Johannes Kepler University Linz, Altenbergerstra{\ss}e 69, A-4040 Linz, Austria}

\author{Christian Schimpf}
\affiliation{Institute of Semiconductor and Solid State Physics, Johannes Kepler University Linz, Altenbergerstra{\ss}e 69, A-4040 Linz, Austria} 

\author{Marcus Reindl}
\affiliation{Institute of Semiconductor and Solid State Physics, Johannes Kepler University Linz, Altenbergerstra{\ss}e 69, A-4040 Linz, Austria} 

\author{Junpeng Lu}
\affiliation{School of Physics, Southeast University, 211189 Nanjing, China} 

\author{Zhenhua Ni}
\affiliation{School of Physics, Southeast University, 211189 Nanjing, China}

\author{Armando Rastelli}
\email{armando.rastelli@jku.at}
\affiliation{Institute of Semiconductor and Solid State Physics, Johannes Kepler University Linz, Altenbergerstra{\ss}e 69, A-4040 Linz, Austria}    

\author{Petr Klenovsk\'{y}}%
    \email{klenovsky@physics.muni.cz}
    \affiliation{Department of Condensed Matter Physics, Faculty of Science, Masaryk University, Kotl\'a\v{r}sk\'a~267/2, 61137~Brno, Czech~Republic}
    \affiliation{Czech Metrology Institute, Okru\v{z}n\'i 31, 63800~Brno, Czech~Republic}

\begin{abstract} 
The optical properties of excitons confined in initially-unstrained GaAs/AlGaAs quantum dots are studied as a function of a variable quasi-uniaxial stress. To allow the validation of state-of-the-art computational tools for describing the optical properties of nanostructures, we determine the quantum dot morphology and the in-plane components of externally induced strain tensor at the quantum dot positions. Based on these \textsl{experimental} 
parameters, we calculate the strain-dependent excitonic emission energy, degree of linear polarization, and fine-structure splitting using a combination of eight-band ${\bf k}\cdot{\bf p}$ formalism with multiparticle corrections using the configuration interaction method. The 
experimental observations are quantitatively well reproduced by our calculations and deviations are discussed.
\end{abstract}


\maketitle

\section{Introduction}
\label{sec:intro}
Sources of quantum-light are attracting considerable attention as one of the key components in quantum networks~\cite{Kimble2008} and optical quantum simulators.~\cite{AspuruGuzik2012} Among others, quantum dots (QDs) are regarded as the best solid-state quantum light emitters.~\cite{Aharonovich2016,Senellart2017,zhou2023epitaxial} Pioneering work on QDs in the field of quantum science and technology has been mostly carried out with InGaAs QDs obtained via the Stranski-Krastanow 
growth method,~\cite{Michler2000,zrenner2002,Santori2002,Akopian2006,pressl2008} which were discovered about three decades ago.~\cite{Leonard1993} 
Since then, progress has been accomplished by improving the material quality to reduce charge noise,~\cite{Kuhlmann2015,Lodahl2022} by integrating QDs in photonic structures,~\cite{Lodahl2015,Senellart2017,Liu2019,Wang2019b,Tomm2021} by tailoring the QD properties via external electric,~\cite{Bennett2010a} magnetic,~\cite{Bayer2002} and elastic fields,~\cite{Oyoko2001,seidl2006effect,Singh2010c,Gong2011e,Martin-Sanchez2018} and by implementing advanced excitation schemes.~\cite{Wang2019b,Sbresny2022} 

In parallel to the experimental work, theoretical models and computational methods have been developed to interpret the observed optical properties of QDs~\cite{Stier2000,brasken2000full,baer2005optical,Bester2006,tomic2009excitonic,Schliwa:09,Mittelstadt2022} and also in the attempt to guide their further development.~\cite{Singh2009} If properly and quantitatively validated, such models could be used to design QDs with tailored properties without the necessity of many resource-intensive growth and characterization runs. In view of the large available parameter space, we are convinced that the availability of such predictive computational tools will be key to enable experiments and applications with increasing complexity. 

Unfortunately, the properties of Stranski-Krastanow QDs generally vary significantly from QD to QD in the same sample, resulting in a large spread in their optical properties.~\cite{Seguin2005,Trotta2013} In turn, this is due to the stochastic nature of the self-assembled growth and to the pronounced material intermixing occurring during QD growth, which is largely driven by strain minimization and leads to disordered structures with complex atomic arrangements.~\cite{Keizer2011,Rastelli2008} On top of the difficulty of accurately determining the structural properties of QDs investigated optically, also the values of the externally applied fields are often affected by large uncertainties, especially in the case of strain fields in deformable few-hundred nm-thick layers (usually referred to as ``membranes''.~\cite{Trotta2012a,Yuan2018a} Although existing computational tools allowed many experimental results to be qualitatively explained,~\cite{Seguin2005,Jons2011,Honig2014,Yuan2018a} the above mentioned uncertainties have hindered their quantitative validation.

Complementary to Stranski-Krastanow QDs, GaAs QDs in AlGaAs nanoholes~\cite{Rastelli2004,Wang2009,Plumhof2010,Plumhof2013,Huo2013a,Yuan2018a,Huang2021a,Heyn2010,Lobl2019} have emerged as a model system to validate electronic-structure and optical properties calculation methods, because of their high ensemble homogeneity,~\cite{DaSilva2021,Keil2017a,Rastelli2004a} negligible built-in strain, and limited intermixing between GaAs core and AlGaAs barriers. In addition, GaAs QDs in nanoholes obtained via Al local-droplet-etching 
on AlGaAs~\cite{Heyn2009,DaSilva2021} have outperformed other optically-active QDs as sources of indistinguishable photons,~\cite{Zhai2022} highly polarization-entangled photon pairs,~\cite{Krapek2010,Huber2018,Schimpf2021} and as hosts of coherent electron spins.~\cite{Zaporski2023} The strongly reduced strain inhomogeneities and alloy disorder compared to Stranski-Krastanow QDs, combined with externally applied stress and electromagnetic fields are also opening the route to the controlled manipulation of nuclear spins in these QDs,~\cite{Chekhovich2020} which -- once carefully optimized -- may act as long-sought semiconductor-based quantum memories.~\cite{taylor2003} 

To allow the validation of state-of-the-art models for describing the optical properties of QDs under stress, we investigate in this work the photoluminescence (PL) spectra of single GaAs/Al$_{0.4}$Ga$_{0.6}$As QDs with well-characterized morphology and subject to variable and experimentally-determined stress configurations. Elastic stress is applied by bonding the QD-containing layer on top of a piezoelectric actuator,~\cite{Ding2010,Kumar2011} and the local stress configuration at the QD positions is determined by using the PL of free excitons in a bulk-like GaAs layer placed right on top of the QD layer (within the same laser excitation spot) and eight-band ${\bf k}\cdot{\bf p}$ theory as local strain gauge.~\cite{Martin-Sanchez2016} The experimentally determined stress and structural parameters are then used as input for eight-band ${\bf k}\cdot{\bf p}$ calculations combined with configuration interaction (CI) to validate the predictions [see Fig.~\ref{fig:Device}~(d) later on] against the measured excitonic emission energy, fine-structure splitting (FSS), and degree of linear polarization (DOLP). The main novelty of this work compared to former combined experimental-theoretical studies is that no assumptions on the local strain configuration at the QD position has been made, allowing us to thoroughly test the validity of the calculation results obtained by state-of-the-art eight-band ${\bf k}\cdot{\bf p}$~\cite{Birner2007} and CI~\cite{Klenovsky2012,Klenovsky2017,Klenovsky2019} with varying number of basis states. In particular, the increase of CI basis states allows us to study the convergence of the method for all studied parameters. We find that the theory results are in good quantitative agreement with the experimental results on exciton emission energy, FSS, polarization direction, and DOLP, the latter being related to heavy hole~(HH)-light~hole~(LH) mixing. Some deviations are nevertheless observed depending on the specific choice of the CI basis states.

The manuscript is structured as follows. Section~\ref{sec:outline} presents the experimental results. In particular, we show the QD morphology, obtained from atomic force microscopy (AFM), a method for determining the local stress configuration produced by a piezoelectric actuator placed below the QD structure, and the effects produced by stress on the emission of neutral excitons confined in QDs for a given strain. That is followed in Sec.~\ref{sec:theorDesc} by detailed description of our theory methods. Thereafter, in Sec.~\ref{sec:expTeorCompar} we present our theory results and compare them with the experiment. Finally, we conclude with a discussion of the open challenges.

\section{Experimental results}
\label{sec:outline}

\begin{figure}[htbp]
	\includegraphics[width=85mm]{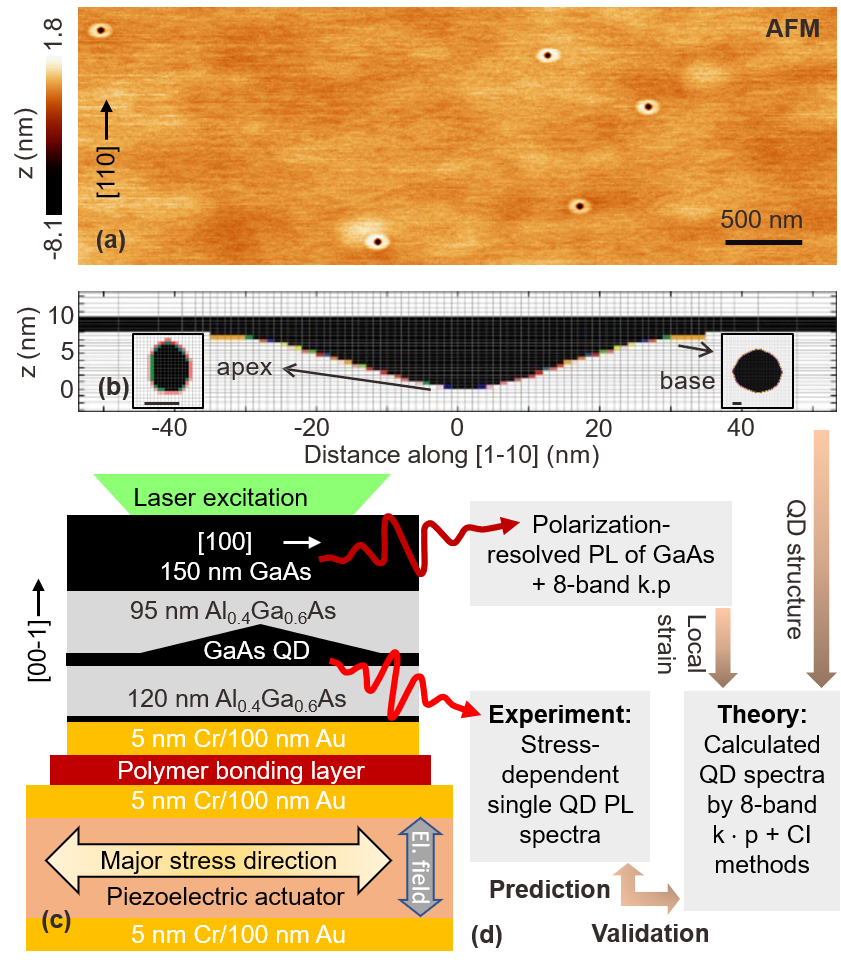}	
	\caption{(a)~Representative AFM topography of an Al$_{0.4}$Ga$_{0.6}$As surface with Al-droplet-etched nanoholes prior to GaAs filling. (b)~Side view of the GaAs QD model used in the calculations, derived from one of the nanoholes. Overgrowth of the Al$_{0.4}$Ga$_{0.6}$As surface with 2~nm GaAs results in nanohole filling and in a quantum well. The insets show cuts parallel to growth plane close to the apex and base of the nanohole with scale bars corresponding to 10~nm. (c)~Cross section of the layer structure of the strain-tunable device used in this work. A semiconductor membrane containing QDs is bonded on top of a piezoelectric actuator. The direction of the major stress axis generated by the piezoelectric actuator is parallel to the non-polar [100] crystal direction of the semiconductor material. Quasi-uniaxial stress can be applied to QDs by applying an electric field between top and bottom sides of the actuator. (d)~Flow diagram of this work, which combines precisely-determined strain and QD structure with theoretical calculations of the experimentally measured QD spectra.}
	\label{fig:Device}
\end{figure} 


%
The experiments were performed on GaAs QDs fabricated by GaAs infilling of nanoholes produced via the local-droplet-etching method on an (001)-oriented AlGaAs surface.~\cite{DaSilva2021, Gurioli2019} As in former works, we argue that the shape of the QDs can be obtained from ex-situ scanning-probe microscopy images of unfilled nanoholes~\cite{Rastelli2004,Wang2009,Huo2014,Huo2017} because of the limited intermixing between GaAs and AlGaAs. This allows us to overcome most of the long-standing obstacles encountered when modeling conventional InGaAs QDs and arising from their complex and poorly known structural properties. Fig.~\ref{fig:Device}~(a) shows an AFM image of representative nanoholes on the sample surface. We see that all nanoholes have similar sizes and shapes. Their average depth (diameter) is 8.4$\pm$0.4~nm (55$\pm$2~nm). In view of the relatively large size of the nanoholes, we expect AFM-tip convolution effects to be negligible. Nevertheless, we used the nominal structural parameters of the used probes (tip radius of 10~nm, tip angle of 18$^{\circ}$) to perform tip-deconvolution using the Gwyddion software.~\cite{Necas2012} The resulting cross-section profile of the GaAs QD implemented in our calculation, which was obtained from the deconvoluted AFM data of an etched nanohole plus a 2~nm-thick quantum well, is shown in Fig.~\ref{fig:Device}~(b). The insets of Fig.~\ref{fig:Device}~(b) show cuts of the structure performed parallel to the growth plane close to the apex and base of the nanohole. The nanohole is slightly elongated along the [110] and [1$\overline{1}$0] crystal directions at its base and apex, respectively. 

To study the properties of GaAs QDs under externally-induced stress, we have fabricated a strain-tunable device, as illustrated in Fig.~\ref{fig:Device}~(c). 
The semiconductor structure consists of the following layer sequence grown by molecular beam epitaxy on a GaAs(001) substrate: an Al-rich AlGaAs-sacrificial layer (used for substrate removal), a 150~nm thick GaAs layer (acting as local strain-gauge), a 95~nm Al$_{0.4}$Ga$_{0.6}$As barrier with nanoholes filled by depositing 2~nm GaAs (QD layer), a 120~nm Al$_{0.4}$Ga$_{0.6}$As barrier, and a final 4~nm GaAs protective layer. To apply variable stress on this structure, we bonded it on a piezoelectric substrate via a flip-chip process. We first coated the QD-containing sample with Cr/Au (5/100~nm) layers acting as a backside mirror to enhance the light extraction efficiency~\cite{Trotta2012a} after flip-chip and cut the sample into small pieces (about 2$\times$2~mm$^2$).
The piezoelectric actuator was derived from a 300~$\mu$m thick, [110]-cut [Pb(Mg$_{1/3}$Nb$_{2/3}$)O$_{3}$]$_{0.72}$[PbTiO$_{3}$]$_{0.28}$ substrate, which we lapped and polished to a final thickness of 100~$\mu$m. After cleaning, top and bottom sides of the actuator were coated with Cr/Au (5/100~nm) electrodes. 
One of the Au-coated sample pieces was then bonded with the piezoelectric actuator using a polymer as adhesion layer~\cite{Martin-Sanchez2018} and by carefully placing its [100] crystal direction aligned along the major stress axis produced by the underlying actuator.
To allow for optical access and increase the range of achievable strain values, the GaAs substrate was then removed via a series of chemical etchings steps, leaving only the optically active structure, bonded on the piezoelectric substrate, as illustrated in Fig.~\ref{fig:Device}~(c).

With the employed configuration, we expect the actuator to produce variable quasi-uniaxial stress along the non-polar [100] crystal direction of (Al)GaAs. (For a discussion on the effect produced by application of stress along the natural cleaving directions of GaAs,~i.e., the [110] and [1$\overline{1}$0], used in previous experiments,~\cite{seidl2006effect,Zhai2020a} see Appendix~I). Both compressive and tensile stress can be continuously transferred to the semiconductor structure by swapping the electric field direction between top and bottom side of the piezoelectric actuator. We note that the stress configuration for a given field applied to the actuator cannot be faithfully predicted because of the poorly known material properties of piezoelectric actuator and bonding material at the measurement temperature. For this reason, strain is often assumed to vary linearly with the electric field in the piezo-material. Since GaAs is a direct bandgap semiconductor with excellent PL and well-studied optical properties under deformation,~\cite{Chuang2012,Vurgaftman2001} the top 150~nm thick GaAs layer was intentionally added to act as a local strain gauge to extract the precise strain configuration at the QD position using the eight-band ${\bf k}\cdot{\bf p}$ method.~\cite{Martin-Sanchez2016}

We now present our stress-depentent PL measurements. The device discussed above was mounted in a He-flow cryostat and all measurements were performed at a temperature of~$\sim$8~K. A laser with wavelength of 532~nm was employed for above-bandgap excitation. Linear-polarization-resolved measurements were performed using a rotatable $\lambda$/2 half-wave plate and a fixed linear polarizer in front of a spectrometer equipped with a CCD camera with a combined spectral resolution of about 40~$\mu$eV.
\begin{figure}[htbp]
	\includegraphics[width=85mm]{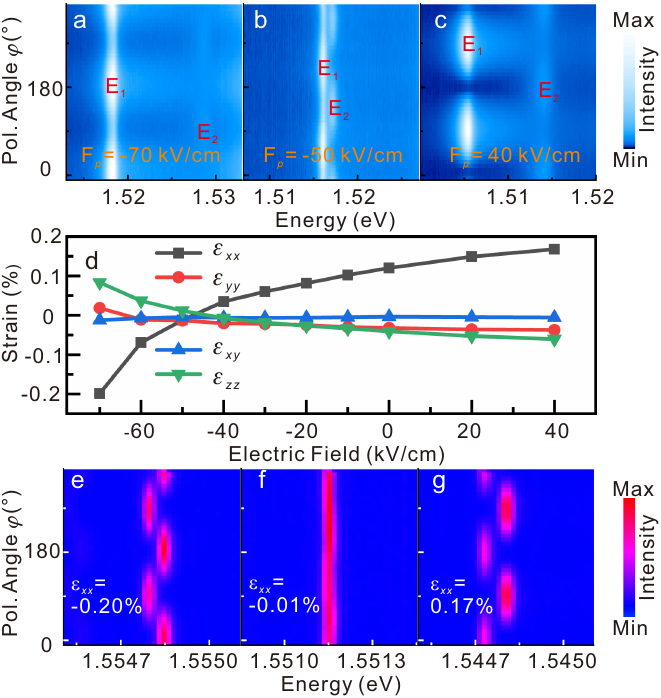}
	\caption{(a),~(b),~and~(c) Representative color-coded linear-polarization-resolved PL spectra of free excitons in a bulk-like GaAs layer under different strain configurations (electric fields applied to the piezoelectric actuator are indicated). E1 and E2 mark the two strain-split emission peaks. Due to the large intensity difference of E1 and E2 under large stress, the logarithm of the measured PL intensity (spanning 4 orders of magnitude) is color-coded for better visibility. (d) Extracted strain components from bulk-like GaAs layer under different electric fields applied to the piezoelectric actuator. Here~$x$,~$y$,~and~$z$ refer to the [100],~[010],~and~[001] crystal orientations of GaAs, respectively. (e),~(f),~and~(g) give the corresponding color-coded linear polarization resolved PL spectra of GaAs QD under the strain configurations of (a),~(b),~and~(c), respectively. The extracted strain tensor component $\epsilon_{xx}$ is marked in each panel. Linear color scale is used for the spectrum intensity.}
	\label{fig:PL}
\end{figure}

Fig.~\ref{fig:PL}~(a),~(b),~and~(c) show color-coded PL spectra of free excitons recombining in the thick GaAs layer under different strain states. Two polarized emission lines (marked as E1 and E2) are clearly observed because an in-plane stress breaks the crystal symmetry and removes the valence band degeneracy at the $\Gamma$ point of the Brillouin zone. In order to determine the configuration of the applied stress, a simple model capable of predicting the optical properties of strained GaAs is used. First, we assume that only the in-plane stress components have finite values since the top membrane surface is free. The strain tensor components corresponding to a certain in-plane stress configuration are included into the Pikus-Bir Hamiltonian for the ${\Gamma}$ point of the first Brillouin zone. That Hamiltonian is thereafter diagonalized in order to obtain single-particle eigenenergies and eigenstates of electrons and holes. The transition probabilities are then calculated within the dipole approximation for different in-plane polarization directions. In general, two emission lines with elliptical polarization are obtained once the strain configuration is defined. 

To perform the inverse operation,~i.e., to find the strain configuration from measured polarization-resolved PL spectra of GaAs [see,~e.g., Fig.~\ref{fig:PL}~(a)-(c)], we take three quantities,~i.e., the energy of the emission lines E1~and~E2,~and the polarization angle ${\phi}_{1}$ of E1 and use a non-linear least-square minimization algorithm to find the in-plane stress configuration which minimizes the deviation between measured and computed parameters. A first guess for this best-fit procedure is made by analytical diagonalization of the Pikus-Bir Hamiltonian without considering the split-off band. 
Using that approach, we extract the strain configurations for different electric fields applied to the piezoelectric actuator, as illustrated in Fig.~\ref{fig:PL}~(d). From the plot we see that the membrane with QDs is under tensile stress at zero applied field, which we attribute to thermally-induced stress during bonding as well as stress induced during the piezoelectric poling process. A close-to-unstrained state is recovered for an applied electric field of about $-50$~kV/cm [Fig.~\ref{fig:PL}~(b)] and we confirm that the applied stress is close to being uniaxial along the [100] ($x$) crystal direction of the semiconductor membrane. In fact, $\epsilon_{xx}$ is by far the largest strain tensor component and it varies between $-0.2$\% and $+0.2$\%, the magnitude of the shear component $\epsilon_{xy}$ is negligible, and the transversal components $\epsilon_{yy}$ and $\epsilon_{zz}$ have similar magnitude. Deviations from a uniaxial stress configurations ($\epsilon_{zz}>\epsilon_{yy}$) are attributed to the fact that the QD-containing layer is biaxially clamped on the underlying piezoelectric actuator. From Fig.~\ref{fig:PL}~(d) we also see that the dependence of the strain components on electric field is strongly non-linear at high fields, highlighting the importance of a direct strain determination for the correct interpretation of data collected on the QDs.

In order to take into account for possible lateral strain inhomogeneities across the membrane, we collect PL spectra of a GaAs QD at the same position as the one used for collecting the bulk-like GaAs emission,~i.e., without displacing the sample with respect to the laser spot. Fig.~\ref{fig:PL}~(e)-(g) shows color-coded linear-polarization-resolved spectra of such a GaAs QD. Since the QD is far away from the edges of the membrane, strain relaxation along its thickness can be excluded and we can safely assume that the strain configurations shown in Fig.~\ref{fig:PL}~(d) coincides with that at the QD location. 
A very small FSS value ($\sim$8.1~$\mu$eV) and an average emission energy of 1.55115~eV are observed for the minimum strain value ($\epsilon_{xx}=-0.01\%$) in Fig.~\ref{fig:PL}~(f), while the FSS value considerably increases with larger strain (both compressive and tensile), as shown in Fig.~\ref{fig:PL}~(e)~and~(g). At the same time the transition energy blue/red shifts under compression/tension and the polarization direction of the high energy component of the excitonic doublet aligns parallel/perpendicular to the [100] direction. 

Besides the spectra and corresponding strain configurations of the QD shown in Fig~\ref{fig:PL}, we performed similar measurements for several other QDs, finding similar results. A further example of stress-dependent QD spectra, the corresponding strain configuration, as well as the strain-dependent optical properties of the neutral exciton can be found in Figs.~\ref{fig:QD2}~and~\ref{fig:ExpQD2Exciton} in Appendix~II.

All these observations qualitatively fit with previous studies.~\cite{Plumhof2011a,Trotta2015a,Yuan2018a,Gong2011e} 
We highlight, however, the importance of determining the strain produced by the actuator when interpreting the experimental data, as discussed more in detail in Sec.~IV. 

\section{Theory model for description of GaAs QD emission}
\label{sec:theorDesc}

Here we describe the theoretical model which we employ in this work. 
We use a combination of the eight-band ${\bf k}\cdot{\bf p}$ method,~\cite{Birner2007,t_zibold,t_klenovsky,t_csontosova,Mittelstadt2022} providing single-particle basis states for the configuration interaction (CI)~\cite{Schliwa:09,Klenovsky2017,Klenovsky2019} algorithm. During the CI calculation our code evaluates also the emission radiative lifetime,~\cite{Klenovsky2017,Klenovsky2019} utilizing the Fermi's golden rule.~\cite{Dirac1927}

In the calculation we first implement the 3D QD model structure (size, shape, chemical composition), see Fig.~\ref{fig:Device}~(b). This is followed by the calculation of elastic strain by minimizing the total strain energy in the structure and subsequent evaluation of piezoelectricity up to nonlinear terms.~\cite{Bester:06,Beya-Wakata2011,Aberl2017,Klenovsky2018} At this point the applied quasi-uniaxial strain tensor, see Fig.~\ref{fig:PL}~(d), is added to the computed one for QD. The resulting strain and polarization fields then enter the eight-band $\mathbf{k}\!\cdot\!\mathbf{p}$ Hamiltonian.

In our ${\bf k}\cdot{\bf p}$, we consider the single-particle states as linear combination of $s$-orbital~like and $x$,~$y$,~$z$~$p$-orbital~like Bloch waves~\cite{Klenovsky2017,Csontosova2020,Mittelstadt2022} at $\Gamma$ point of the Brillouin zone,~i.e.,
\begin{equation}
    \Psi_{a_n}(\mathbf{r}) = \sum_{\nu\in\{s,x,y,z\}\otimes \{\uparrow,\downarrow\}} \chi_{a_n,\nu}(\mathbf{r})u^{\Gamma}_{\nu}\,,
\end{equation}
where $u^{\Gamma}_{\nu}$ is the Bloch wave-function of $s$- and $p$-like conduction and valence bands at $\Gamma$ point, respectively, $\uparrow$/$\downarrow$ marks the spin, and $\chi_{a_n,\nu}$ is~the~envelope function for $a_n \in \{ e_n, h_n \}$ [$e$ ($h$) refers to electron (hole)] of the $n$-th single-particle state.
Thereafter, the following envelope-function $\mathbf{k}\!\cdot\!\mathbf{p}$ Schr\"{o}dinger equation is solved
\begin{equation}
\label{eq:EAkp}
\begin{split}    
    &\sum_{\nu\in\{s,x,y,z\}\otimes \{\uparrow,\downarrow\}}\left(\left[E_\nu^{\Gamma}-\frac{\hbar^2{\bf \nabla}^2}{2m_e}+V_{0}({\bf r})\right]\delta_{\nu'\nu}+\frac{{\hbar}{\bf \nabla}\cdot{\bf p}_{\nu'\nu}}{m_e}\right.\\
    &\left. + \hat{H}^{\rm str}_{\nu'\nu}({\bf r})+\hat{H}^{\rm so}_{\nu'\nu}({\bf r})\right)\chi_{a_n,\nu}({\bf r})=E^{k\cdot p}_n\cdot \chi_{a_n,\nu'}({\bf r}),
\end{split}    
\end{equation}
where the term in round brackets on left side of the equation is the envelope function $\mathbf{k}\!\cdot\!\mathbf{p}$ Hamiltonian $\hat{H}_0^{k\cdot p}$, and $E^{k\cdot p}_n$ on the right side is the $n$-th single-particle eigenenergy. Furthermore, $E_\nu^{\Gamma}$ is the energy of bulk $\Gamma$-point Bloch band $\nu$, $V_0({\bf r})$ is the scalar potential (e.g. due to piezoelectricity), $\hat{H}^{\rm str}_{\nu'\nu}({\bf r})$ is the Pikus-Bir Hamiltoninan introducing the effect of elastic strain,~\cite{Birner2007,t_birner,t_zibold,t_klenovsky,t_csontosova} and $\hat{H}^{\rm so}_{\nu'\nu}({\bf r})$ is the spin-orbit Hamiltonian.~\cite{t_birner,t_zibold,t_klenovsky,t_csontosova} Further, $\hbar$ is the reduced Planck's constant, $m_e$ the free electron mass, $\delta$ the Kronecker delta, and $\nabla := \left( \frac{\partial}{\partial x}, \frac{\partial}{\partial y}, \frac{\partial}{\partial z} \right)^T$.

We use single-particle states computed by the aforementioned ${\bf k}\cdot{\bf p}$ as basis states for our CI. In CI we consider the excitonic (X) states as linear combinations of the Slater determinants
\begin{equation}
    \psi_i^{\rm X}(\mathbf{r}) = \sum_{\mathit m=1}^{n_{\rm SD}} \mathit \eta_{i,m} \left|D_m^{\rm X}\right>, \label{eq:CIwfSD}
\end{equation}
where $n_{\rm SD}$ is the number of Slater determinants $\left|D_m^{\rm X}\right>$, and $\eta_{i,m}$ is the $i$-th CI coefficient which is found along with the eigenenergy using the variational method by solving the Schr\"{o}dinger equation 
\begin{equation}
\label{CISchrEq}
\hat{H}^{\rm{X}} \psi_i^{\rm X}(\mathbf{r}) = E_i^{\rm{X}} \psi_i^{\rm X}(\mathbf{r}),
\end{equation}
where $E_i^{\rm{X}}$ is the $i$-th eigenenergy of excitonic state $\psi_i^{\rm X}(\mathbf{r})$, and~$\hat{H}^{\rm{X}}$ is the CI Hamiltonian which reads
\begin{equation}
\label{CIHamiltonian}
\hat{H}^{\rm{X}}=\hat{H}_0^{k\cdot p}+\hat{V}^{\rm{X}},
\end{equation}
with $\hat{H}_0^{k\cdot p}$ and $\hat{V}^{\rm{X}}$ representing the Hamiltonian of the noninteracting single-particle states, defined above, and the~Coulomb interaction between Slater determinants constructed from those for exciton, respectively. Further, the matrix element of $\hat{V}^{\rm{X}}$ in basis of the Slater determinants $\left|D_m^{\rm X}\right>$ is~\cite{Klenovsky2017,Klenovsky2019,Csontosova2020}
\begin{equation}
\begin{split}
    &V^{\rm X}_{n,m}=\bra{D_n^{\rm X}}\hat{J}\ket{D_m^{\rm X}} = -\frac{1}{4\pi\epsilon_0} \sum_{ijkl} \iint {\rm d}\mathbf{r} {\rm d}\mathbf{r}^{\prime} \frac{e^2}{\epsilon(\mathbf{r},\mathbf{r}^{\prime})|\mathbf{r}-\mathbf{r}^{\prime}|} \\
    &\times \{ \Psi^*_i(\mathbf{r})\Psi^*_j(\mathbf{r}^{\prime})\Psi_k(\mathbf{r})\Psi_l(\mathbf{r}^{\prime}) - \Psi^*_i(\mathbf{r})\Psi^*_j(\mathbf{r}^{\prime})\Psi_l(\mathbf{r})\Psi_k(\mathbf{r}^{\prime})\}
    \\
    &= \sum_{ijkl}\left(V^{\rm X}_{ij,kl} - V^{\rm X}_{ij,lk}\right).\\
\end{split}
\label{eq:CoulombMatrElem}
\end{equation}
Here $\hat{J}$ marks the Coulomb operator, $e$ labels the elementary charge and $\epsilon(\mathbf{r},\mathbf{r}^{\prime})$ is the spatially dependent relative dielectric function, $\epsilon_0$ is the vacuum permittivity. Note that for $\epsilon(\mathbf{r},\mathbf{r}^{\prime})$ in Eq.~\eqref{eq:CoulombMatrElem} we use the position-dependent bulk dielectric constant. The Coulomb interaction in Eq.~\eqref{eq:CoulombMatrElem} described by $V^{\rm X}_{ij,kl}$ ($V^{\rm X}_{ij,lk}$) is called direct (exchange). Further, the multipole expansion of the exchange interaction is included in our CI following the theory outlined in Refs.~\onlinecite{Takagahara2000,Krapek2015,klenovsky2022interplay}.

The sixfold integral in Eq.~\eqref{eq:CoulombMatrElem} is evaluated using the~Green's function method.~\cite{Schliwa:09,Stier2000,Klenovsky2017,Csontosova2020} The integral in Eq.~\eqref{eq:CoulombMatrElem} is split into solution of the Poisson's equation for one quasiparticle $a$ only, followed by a three-fold integral for quasiparticle $b$ in the electrostatic potential generated by particle $a$ and resulting from the previous step. That procedure, thus, makes the whole solution numerically more feasible and is described by
\begin{equation}
\begin{split}
    \nabla \left[ \epsilon(\mathbf{r}) \nabla \hat{U}_{ajl}(\mathbf{r}) \right] &= \frac{4\pi e^2}{\epsilon_0}\Psi^*_{aj}(\mathbf{r})\Psi_{al}(\mathbf{r}),\\
    V^{\rm{X}}_{ij,kl} &= \int {\rm d}\mathbf{r}'\,\hat{U}_{ajl}(\mathbf{r}')\Psi^*_{bi}(\mathbf{r}')\Psi_{bk}(\mathbf{r}'),
\end{split}
\label{eq:GreenPoisson}
\end{equation}
where $a,b \in \{e,h\}$.

During computation of CI, the oscillator strength of the exciton optical transition $F^{\rm X}_{fi}$ between the $i$-th and $f$-th eigenstate of the excitonic complex, respectively, is evaluated using the Fermi's golden rule~\cite{Zielinski2010,Klenovsky2015}
\begin{equation}
\label{eq:CIOscStrength}
F^{\rm X}_{fi}=\left|\left<{\rm X}_f\left|\hat{P}\right|{\rm X}_i\right>\right|^2,
\end{equation}
where $\left|{\rm X}_f\right>$ is the final state after recombination of electron-hole pair in $\left|{\rm X}_i\right>$. The operator $\hat{P}$ is defined by
\begin{equation}
\label{eq:CIOscStrengthSP}
\hat{P}=\sum_{rp}\left<\Psi_{e_r}\left|{\bf e}\cdot\hat{{\bf p}}\right|\Psi_{h_p}\right>.
\end{equation}
Here $\Psi_e$ and $\Psi_h$ are the single-particle wavefunctions for electrons and holes, respectively, ${\bf e}$ is the polarization vector, and $\hat{{\bf p}}$ is the momentum operator.

\section{Comparison between theory and experiment for strained GaAs QDs}
\label{sec:expTeorCompar}


As discussed above, we focus on the effect of variable quasi-uniaxial stress on several properties of the ground state exciton (interacting electron-hole pair), namely the emission energy, FSS value, and DOLP. In view of the fact that the FSS has a rather feeble magnitude, it is important to reduce the influence of numerical errors in our computations as much as possible, which we outline in the following. 

First, we notice that the CI produces a correction to the transitions between electron and hole ground states by introducing an overall energy shift due to the Coulomb attraction and correlation and by splitting the single transition among single-particle states into four excitonic states via the exchange interaction. 
Furthermore, the energy difference between the single-particle electron ground state and first excited states is larger than that for holes, namely in the case of our QDs $17$~meV and $2$~meV for the former and latter, respectively. This difference is due to the different effective mass of electrons and heavy holes ($0.067\,m_e$ and $0.51\,m_e$ for GaAs~\cite{Vurgaftman2001}, where $m_e$ is free electron mass), the latter representing 94\% in our single-particle ground-state hole wavefunctions for the case of absence of external stress.
Because of the closely spaced hole levels compared to the electron levels, we take as our CI basis the single-particle electron ground state doublet and the single-particle hole states up to an energy of about 15~meV about the ground state. Using this procedure we numerically speed-up our calculations compared to the case in which also the number of electron single-particle states is increased and also omit integrals between excited single-particle electron and excited single-particle hole states, which in turn limits the amount of possible numerical errors.~\footnote{Excited states with increasing energy have increasing number of nodes and stronger spatial variations. Numerical integrals between them are thus increasingly prone to errors because of the finite size of the spatial grid.}

\begin{figure}[htbp]
	\centering
    \includegraphics[width=85mm]{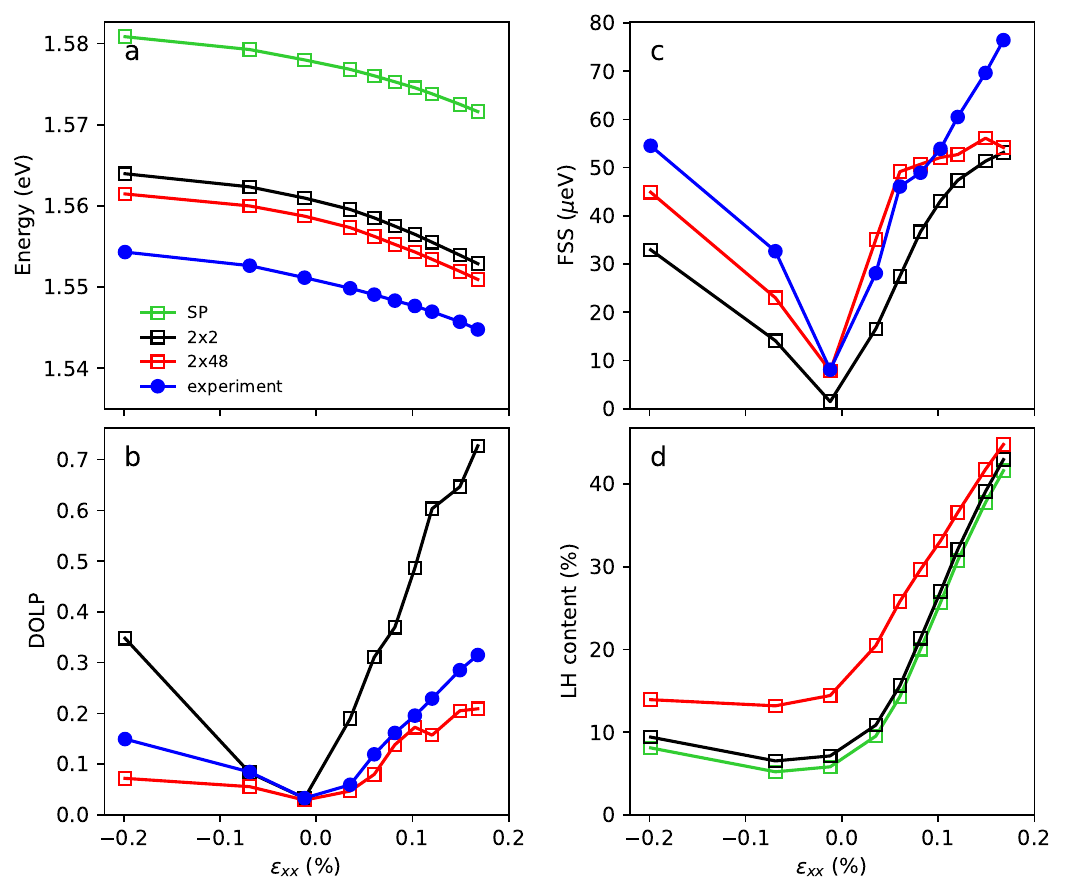}
	\caption{Comparison between theory and experiment for a strained GaAs QD.
	Measured and calculated (a)~energy of the neutral exciton X, (b)~degree of linear polarization, and (c)~FSS value as a function of the externally induced strain $\epsilon_{xx}$. (d)~Light hole content in the exciton vs. $\epsilon_{xx}$. The values of the other components of the strain tensor are provided in Fig.~\ref{fig:PL}~(d) and the experimental data are extracted from PL spectra as those shown in Fig.~\ref{fig:PL}~(e)--(g). In panels (a)--(c) we show the experiment (blue circles) and in (a)--(d) calculations performed without considering Coulomb interaction between electrons and holes (green squares) and those obtained using CI with single-particle bases of $2\times2$ (black squares) and $2\times48$ (red squares) electron~$\times$~hole single-particle states.
 }
	\label{fig:ExpVsCI}
\end{figure}

The calculation results are presented in Fig.~\ref{fig:ExpVsCI} together with the experimental results, which are extracted from polarization-resolved QD spectra as those shown in Fig.~\ref{fig:PL}~(e)-(g). In Fig.~\ref{fig:ExpVsCI}~(a) we see that the increase of applied tension causes a reduction of mean exciton energy both in experiment and theory by a similar magnitude. We also observe differences for theory results obtained for different number of CI basis states considered. While considering the Coulomb interaction between electron and hole causes an overall energy reduction of the exciton energy by $\sim17$~meV compared to single-particle transition, an inclusion of the effect of Coulomb correlation further reduces the energy by 
$\sim2$~meV. The residual discrepancy of $\sim5$~meV between theory and experiment is fully compatible with the inhomogeneous broadening of the QD ensemble arising from the observed fluctuations in QD sizes (a QD height increase by 0.5~nm leads to an emission energy shift of about 5~meV). 

We now turn our attention to the excitonic fine structure. While in absence of strain the exciton has dominant HH character [see Fig.~\ref{fig:ExpVsCI}~(d)] and it mostly consists of two bright transition dipoles with polarization in the growth plane, anisotropic in-plane strain makes one of the two initially dark states bright. Since we expect the corresponding transition dipole to be parallel to the growth direction, it is not visible in our experiment, but was observed in Ref.~\onlinecite{Yuan2018a}. We will, thus, focus on the in-plane polarized transition dipoles only, which are split by the FSS, and refer to Fig.~\ref{fig:Teo4X} in Appendix~III for the calculated transition energies of all four excitonic lines.

Fig.~\ref{fig:ExpVsCI}~(b) shows the results for the DOLP of the exciton emission. Here, for all sizes of the CI basis, our theory correctly reproduces the DOLP minimum for $\epsilon_{xx}\simeq 0$, and the magnitude of that approaching zero as in experiment. However, while results obtained by our theory considering Coulomb interaction only and no correlation (black squares)
overestimate the exciton DOLP, 
the Coulomb correlation causes an overall reduction of the DOLP (red squares) and an improvement of the agreement with the experimental values. Some deviations are nevertheless observed for large strain values.

In Fig.~\ref{fig:ExpVsCI}~(c) we plot the results for the values of the FSS between the two bright exciton states. Here, again, the minimum of FSS is reached for $\epsilon_{xx}\simeq 0$ and the magnitude, as well as the change in FSS with strain are similar to experiment. Similarly as for DOLP, the agreement between theory and experiment improves when Coulomb correlation is considered. However, deviations are still present when the tensile strain is increased.
We note that we have repeated the FSS calculations including also the multipole expansion~\cite{Krapek2015,klenovsky2022interplay} of the exchange interaction and considering the nonlinear piezoelectricity~\cite{Bester:06,Beya-Wakata2011,Aberl2017,Klenovsky2018} and both of the aforementioned changed the result within $\sim 2~\rm{\mu eV}$ from the values shown in Fig.~\ref{fig:ExpVsCI}~(c) and, thus, have negligible effect on the discussed results.

\begin{figure}[htbp]
	\centering    
    \includegraphics[width=85mm]{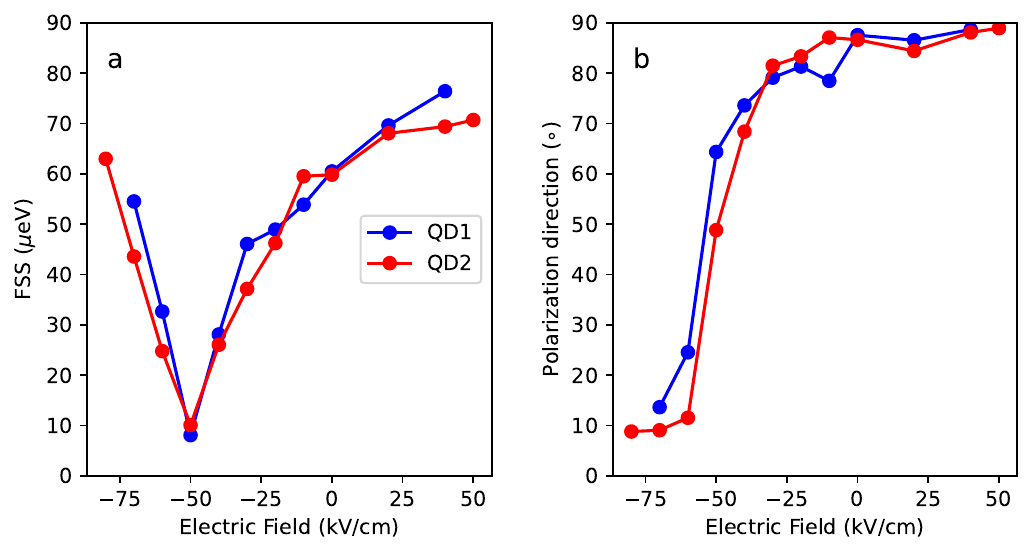}
	\caption{Impact of non-linear dependence of strain on electric field applied to the piezoelectric actuator. (a) Dependence of FSS and (b) orientation of the excitonic high-energy component with respect to the [100] crystal direction for the neutral exciton confined in two QDs vs electric field applied to the piezoelectric actuator. QD1 is the same QD as that discussed in Fig.~\ref{fig:ExpVsCI}, QD2 is a second QD, with corresponding data shown in Appendix~II.
 }
	\label{fig:TeoFSS_V_exx}
\end{figure}

To highlight the importance of the experimental determination of the strain induced by the piezoelectric actuator, we show in Fig.~\ref{fig:TeoFSS_V_exx}~(a) the FSS measurements versus the electric field applied to the piezoelectric actuator for the QD discussed in Fig.~\ref{fig:PL} and also for another QD (see  Fig.~\ref{fig:QD2} of Appendix~II). The corresponding in-plane polarization direction of the high-energy component of the bright exciton is shown in Fig.~\ref{fig:TeoFSS_V_exx}~(b). The plots are still in qualitative agreement with the expected level anticrossing under in-plane anisotropic stress.~\cite{Gong2011e} However, if we would assume a linear dependence of strain on applied electric field as in former works,~\cite{Plumhof2011a,Plumhof2013,Yuan2018a} we would argue a strongly asymmetric behavior of the FSS versus stress. The measured non-linear dependence of $\epsilon_{xx}$ on applied electric field is thus crucial in order to conscientiously test any predictive theory model of stress-tuned QDs.
%


\begin{figure}[htbp]
	\centering
    \includegraphics[width=85mm]{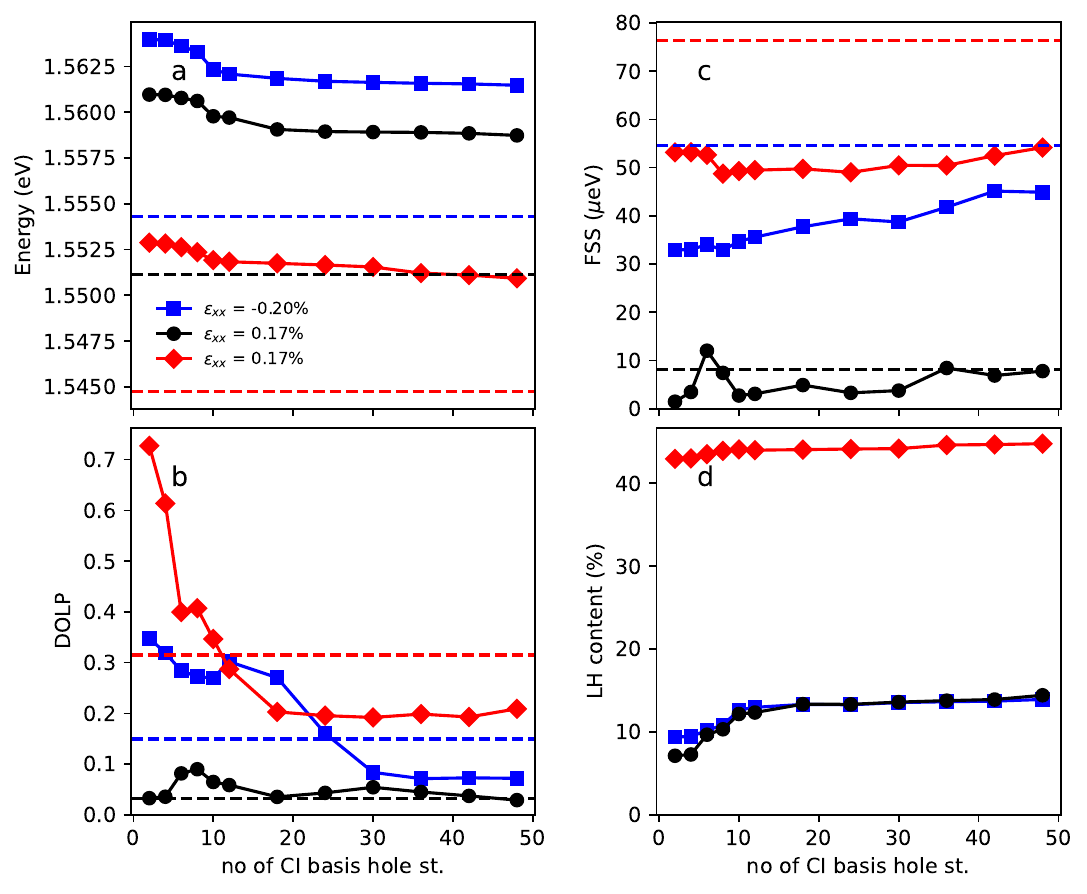}
	\caption{(a)--(d) Dependence of our ${\bf k}\cdot{\bf p}$+CI calculations results on the number of single-particle hole states included in the CI basis for three different configurations with the indicated values of $\epsilon_{xx}$. (Only the two ground-state single-particle electron states are included, see main text). The broken horizontal lines in (a)--(d) show the corresponding experimental values.}
	\label{fig:ExpVsCIconv}
\end{figure}

As in the experiment [see Fig.~\ref{fig:PL}~(e-g) and Fig.~\ref{fig:TeoFSS_V_exx}~(b)], the calculations reproduce the 90$^\circ$ rotation of the polarization direction of the two bright exciton components with varying strain, independent of the used number of CI states (not shown). 

Both the increase in FSS and DOLP can be attributed to the in-plane symmetry breaking introduced by the quasi-uniaxial stress, which in turn affects the valence band structure, modifying the mixing between heavy-hole (HH) and light-hole (LH) bands (and, to a smaller extent, the split-off band, especially under compressive strain). Fig.~\ref{fig:ExpVsCI}~(d) displays the calculated LH content of the ground-state exciton (with 0\% corresponding to a purely HH exciton). We see that the LH content especially increases under tensile strain because the LH and HH states get closer under this condition. In addition, Coulomb correlation effects tend to increase the LH content of the exciton because the excited single-particle hole  states are substantially mixed, as known from previous studies.~\cite{Huo2014}

Since the Coulomb correlation plays a major role in large GaAs QDs,~\cite{Huber2019,Huang2021a} we studied the dependence of the aforementioned exciton properties on CI basis size, increasing the number of single-particle hole states in the CI basis while keeping fixed only the number of electron single-particle ground states as discussed above. These results are shown in Fig.~\ref{fig:ExpVsCIconv}~(a)--(d) for three distinct and representative values of $\epsilon_{xx}$. We see that all four studied parameters of ground state exciton vary little when increasing the number of single-particle hole states included in CI basis beyond about 30. However we see that the FSS keeps increasing with increasing basis size for large strain, possibly further approaching the experimental data. In order to further test the influence of our assumption of computing the properties of ground state exciton with basis formed only from single-particle electron ground states and variable number of single-particle hole states, we have repeated the same calculation for a CI basis formed from the ground state single-particle hole and a variable number of single-particle electron states. The results, which are depicted in Figs.~\ref{fig:ExpVsCIel}~and~\ref{fig:ExpVsCIconvel} of Appendix~III, are very similar to those given in Figs.~\ref{fig:ExpVsCI}~and~\ref{fig:ExpVsCIconv}, respectively, showing a weak dependence on the CI basis choice, as found previously also in Ref.~\onlinecite{Schliwa:09}. However, we note that the agreement between calculation results and experiment for the FSS (DOLP) improves (deteriorates) for the alternative choice of CI basis.
%

\section{Discussion and conclusions}
In this work we have studied GaAs/AlGaAs quantum dots externally tuned by quasi-uniaxial stress. To produce data suitable for the quantitative validation of state-of-the-art computational tools (and different from previous studies), we have determined not only the structural properties of the QDs but also the induced strain configuration at the QD position. 

The experimentally determined QD morphology and strain tensor were thereafter used as input for eight-band ${\bf k}\cdot{\bf p}$ calculations combined with multiparticle corrections using the configuration interaction (CI) method to compute the expectation values of energy, fine-structure splitting, and degree of linear polarization (DOLP) of neutral excitons confined in the QDs. Our calculations reproduce well our experimental results. In particular, the calculated values of FSS are closer to the experiment compared to former works.~\cite{Yuan2018a}

Residual discrepancies between calculations and experiment (of the order of about 30\% for FSS and DOLP) are however still observed. In an attempt to improve the agreement, we have systematically explored the effect of the CI basis size on the numerical results by fixing the CI basis for one quasiparticle to single-particle ground state and varying the number of single-particle states of the other quasiparticle. We observed  convergence in some parameters and slow changes in other parameters, independent on whether the CI basis was fixed for single-particle electrons and the number of holes was varied or vice versa.

As a final remark, we mention that we have also attempted to reproduce the behavior of singly charged excitons confined in GaAs QDs under the effect of stress. While the energetic position of the converged positive trion X$^+$ relative to the neutral exciton X matches rather well the experimental results in absence of strain (relative binding energy $E_B$ of 2.25~meV here versus 3.5~meV$-$4.5~meV in Ref.~\onlinecite{Huang2021a}), the predicted effect of tensile strain on $E_B$ is substantially larger than in the experiment. For the negative trion X$^-$, CI calculations showed poor convergence independent of the used basis and, different from the experiment, resulted in smaller $E_B$ than the value for the X$^+$.

Because of the large number of bound single-particle states obtained by the used ${\bf k}\cdot{\bf p}$ solver in the studied QDs, further testing whether converged results are achievable with our methods requires computational resources out of our current capabilities.

On one hand, this work shows that ${\bf k}\cdot{\bf p}$ combined with the configuration interaction method represents an adequate toolbox to describe experimental results also for large (here about 9~nm tall and 70~nm wide) GaAs nanostructures, in line with previous reports,~\cite{Huo2014, Huang2021a,Huo2017} and the fidelity of that reaches predictive level. On the other hand, we highlight the importance of exploring the effects of basis size in the configuration interaction method, since conclusions on whether calculations quantitatively agree with the experiment may critically depend on that choice. The same considerations most probably apply also to atomistic models~\cite{Huo2013a, Yuan2018a, Heyn2022} used to evaluate the single-particle states needed for configuration interaction.

Once thoroughly validated computational tools will become available, we can imagine using them in combination with advanced search algorithms to retrieve the QD structure and/or the external fields experienced by a QD taking as an input the many lines observed in QD spectra, as envisioned in Ref.~\onlinecite{Mlinar2009}.


To allow other research groups to test the reliability of their computational tools, we make all data presented in this work available under Creative Commons Attribution 4.0 International.~\cite{yuan_xueyong_2023_7748664}

\section{Acknowledgements}
\label{sec:conclusions}
%
The authors thank A. Haliovic and U. Kainz for technical assistance.
X.Y. acknowledges the support by the National Natural Science Foundation of China (NSFC 12104090), “the Fundamental Research Funds for the Central Universities”, and “Zhishan” Scholars Programs (3207022204A2) of Southeast University.

D.C. and P.K. were  financed by the project CUSPIDOR, which has received funding from the QuantERA ERA-NET Cofund in Quantum Technologies implemented within the European Union's Horizon 2020 Programme. In addition, this project has received funding from the Ministry of Education, Youth and Sports of the Czech Republic and from European Union's Horizon 2020 research and innovation framework programme under Grant agreement No. 731473.
The work was also partially funded by projects 20IND05 QADeT, 20FUN05 SEQUME, 17FUN06 SIQUST that received funding from the EMPIR programme co-financed by the Participating States and from the European Union’s Horizon 2020 research and innovation programme.

A.R. acknowledges the Austrian Science Fund (FWF) via the Research Group FG5 and the project I~4320, the Linz Institute of Technology, the QuantERA II Programme that has received funding from the European Union’s Horizon 2020 research and innovation programme under Grant Agreement No 101017733, and via the Austrian Research Promotion Agency (FFG), Grant no. 891366 and the European Union’s Horizon 2020 research and innovation program under Grant Agreements No. 899814 (Qurope) and No. 871130 (Ascent+).


%

\section*{Appendix I. Effect of strain-induced piezoelectricity on QD emission}
\label{appendixI}
Since (Al)GaAs is a piezoelectric material, we need to consider the influence of the piezoelectric effect caused by strain on the optical properties of quantum dots. Electric polarization will be generated inside of (Al)GaAs under shear strain.

\begin{figure}[b!]
    \includegraphics[width=85mm]{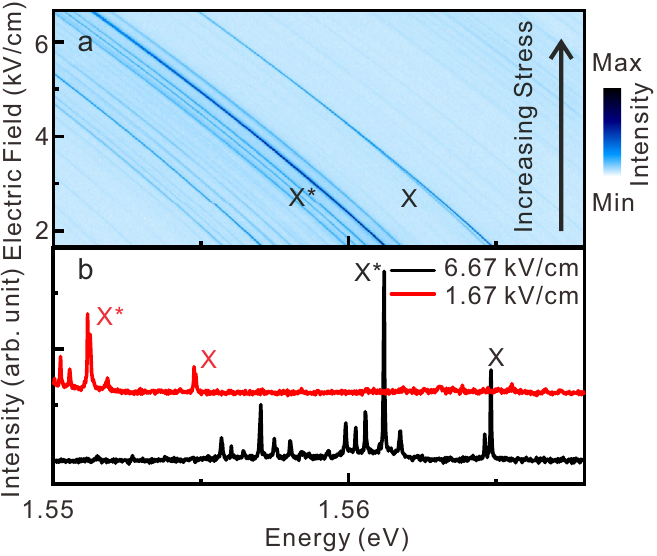}
    \caption{(a) Color coded PL spectra of GaAs QDs emission as a function of the electric field applied to the piezoelectric actuator, the major stress axis is along [110] crystal  direction of AlGaAs. (b) Two typical PL spectra of GaAs QDs for two different stress configurations. X, X$^*$ represent neutral exciton and positively charge exciton, respectively. Linear color scale is used for the spectrum intensity. }
    \label{fig:110}
\end{figure}
If only in-plane uniaxial stress is taken into consideration, two typical cases can be found: if uniaxial stress is along [100] direction, there is no shear strain, which is close to the experimental conditions used to obtain the data presented here. Hence, no electric field is generated and no substantial deterioration of QD optical properties occurs if uniaxial stress is applied along [100] direction. 

However, when the uniaxial stress is applied along the [110] direction, shear strain ($\epsilon_{xy}$) and associated electric field ($E_{[001]}$) read:
\begin{equation}
E_{[001]}=-\frac{e_{14}\epsilon_{xy}}{\varepsilon_{0}(1+\chi )},
\end{equation}
where $e_{14}$ is the piezoelectric tensor coefficient, $\varepsilon_{0}$ the permittivity of free space, and $\chi$ the low frequency dielectric susceptibility. Thus, in the case of uniaxial stress applied along [110] direction, an additional vertical electrical field needs to be taken into account. Two effects can be induced by the vertical electric field: (i) the quantum-confined Stark effect due to induced valence and conduction bands tilting and (ii) the carrier ionization. When the tunneling rate becomes comparable to the radiative recombination rate, the emission peaks get broadened and quenched.

\begin{figure}[!b]
    \includegraphics[width=85mm]{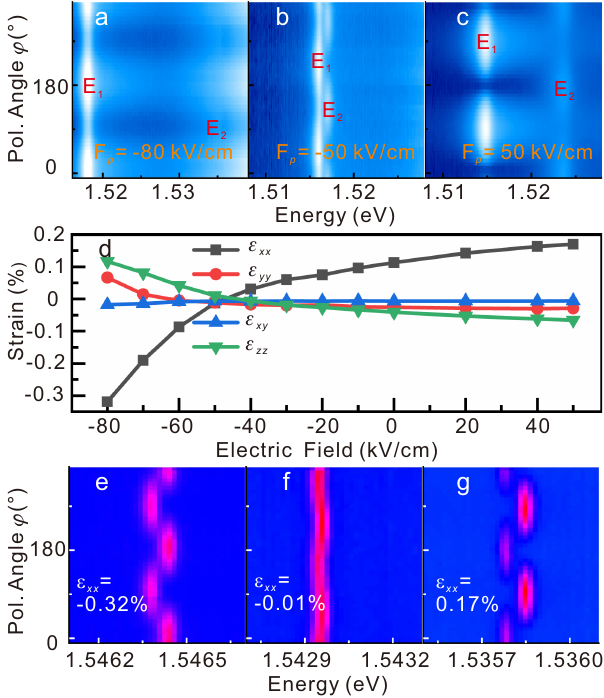}
    \caption{Same as Fig.~\ref{fig:PL}, but for a larger GaAs QD at a different location of the strain-tunable device.}
	\label{fig:QD2}
\end{figure}
We have performed another study to investigate the behavior of QDs under quasi-uniaxial stress along [110] direction. Fig.~\ref{fig:110}~(a) shows PL spectra of a GaAs QD under increasing stress. It is seen that the intensities of emission peaks decrease with increasing stress while their linewidths increase with that. This phenomenon is more evident when emission spectra of QD under different uniaxial stress of [110] direction are compared, as illustrated in Fig.~\ref{fig:110}~(b). The full width at half maximum of the neutral exciton changes from 40~$\mu$eV (resolution limit) to 83~$\mu$eV and that of the charged exciton X$^*$ changes from 51~$\mu$eV to 208~$\mu$eV. All these observations are consistent with our expectations.
\begin{figure}[htbp]
	\centering
    \includegraphics[width=85mm]{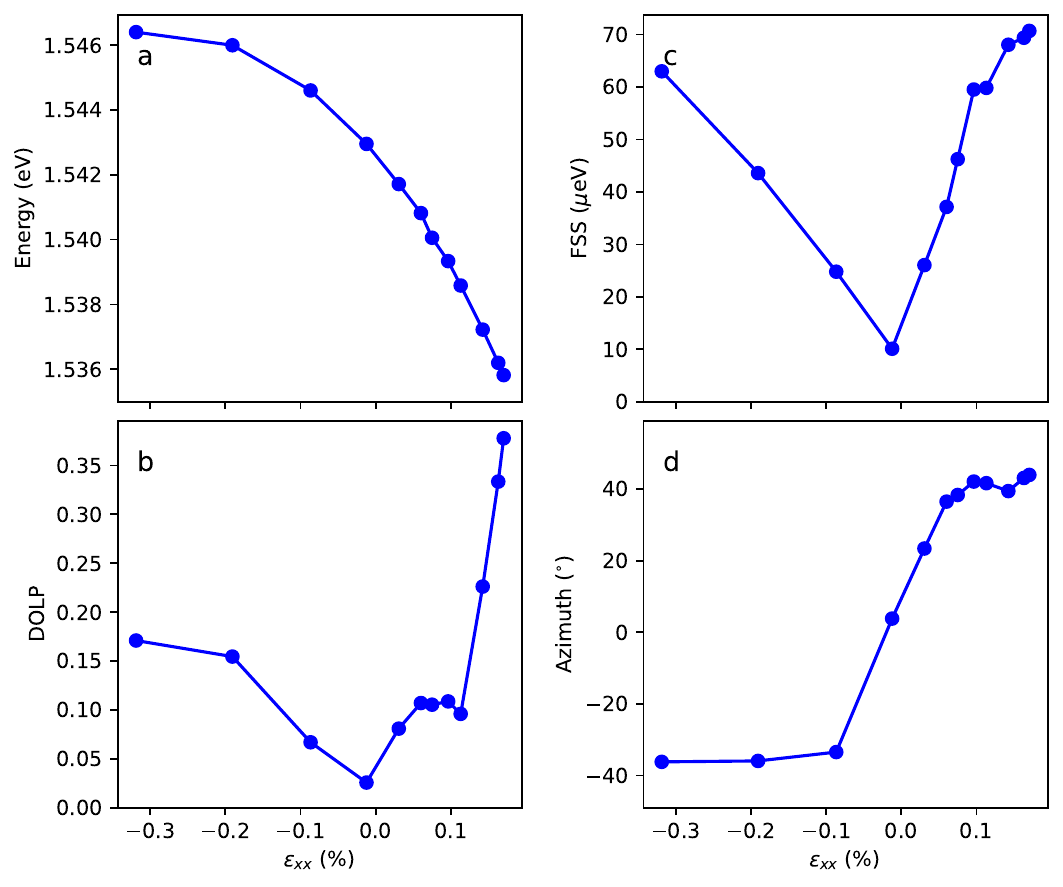}
	\caption{
	Measured (a)~energy of the neutral exciton X, (b)~degree of linear polarization, (c)~FSS value, and (d)~polarization azimuth of exciton as a function of the externally induced strain $\epsilon_{xx}$ from data in Fig.~\ref{fig:QD2}.}
	\label{fig:ExpQD2Exciton}
\end{figure}
In addition, we have occasionally observed a sudden disappearance of PL at certain fields and its reappearance at a slightly different field when ramping back the field. This behavior, especially prominent in devices featuring strain amplification as in Ref.~\onlinecite{Yuan2018a} would need further investigation and shows that stress along polar directions must be kept limited to preserve high optical quality.

\section*{Appendix II. Additional data for a second QD under quasi-uniaxial stress.}
\label{appendixII}
Similar to Fig.~\ref{fig:PL}, Fig.~\ref{fig:QD2} shows the results of the determination of the strain configuration induced by the piezoelectric actuator at the location of another QD, and the corresponding effect on the PL spectrum of the QD. From the PL spectra of the neutral exciton confined in this QD, we extract the data shown in Fig.~\ref{fig:ExpQD2Exciton}. For the unstrained configuration, this QD has an emission energy about 10~meV below that of the QD presented in Fig.~\ref{fig:PL} and was selected on the low energy tail on the QD distribution. In this case the largest strain component $\epsilon_{xx}$ is varied by 0.5\%, corresponding to an energy shift of more than 10~meV for the QD, see Fig.~\ref{fig:ExpQD2Exciton}~(a). In spite of the larger size, the strain-induced effects are similar: a non-linear shift of the emission energy with strain (due to the interaction of HH and LH bands), an anticrossing of the bright excitonic states [see Fig.~\ref{fig:ExpQD2Exciton}~(b)] accompanied by a rotation of the polarization direction of the two excitonic lines by 90$^\circ$ [see Fig.~\ref{fig:ExpQD2Exciton}~(d)], and polarized emission under strain [see Fig.~\ref{fig:ExpQD2Exciton}~(c)]. As in the case of the QD shown in Fig.~\ref{fig:PL}, we see that the DOLP [cf. Fig.~\ref{fig:ExpVsCI}~(c)] shows an asymmetric behavior around the unstrained configuration and is particularly pronounced for tensile strain due to the increased HH-LH mixing [see Fig.~\ref{fig:ExpVsCI}~(d)].

\begin{figure}[!t]
	\centering    
    \includegraphics[width=85mm]{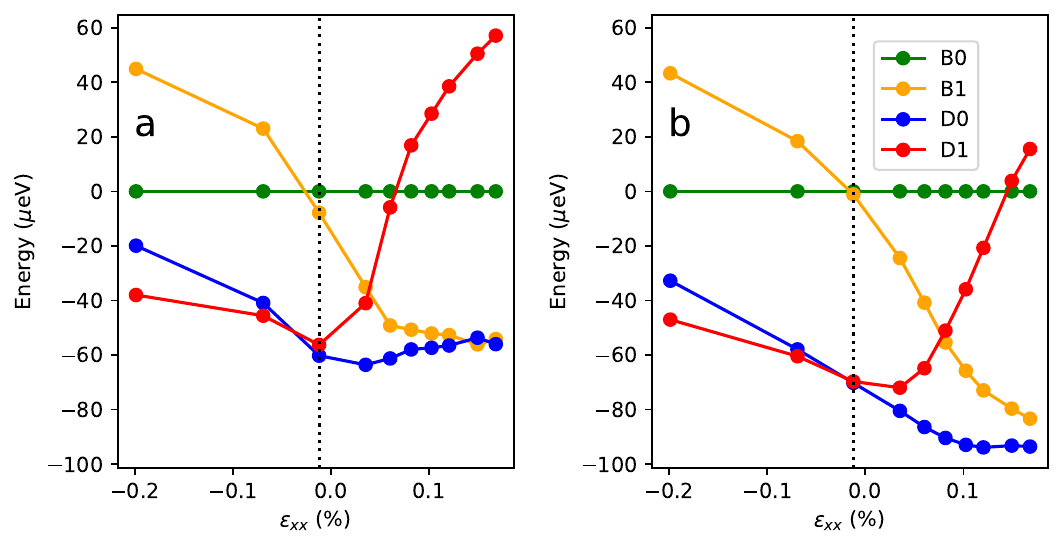}
	\caption{Dependence of the four components of the ground state exciton on $\epsilon_{xx}$. The bright components with in-plane polarized transition dipoles, marked as B0 and B1, are given as green and yellow curves, respectively, while dark and vertically-polarized components, D0 and D1, are given by blue and red curves, respectively. For clarity, the energies here are offset to that of the B0 bright component. Notice that D1 component swaps its energy position with both bright components B0 and B1 for tensile $\epsilon_{xx}$.~\cite{Yuan2018a} The calculations in (a) were performed for 48 single-particle hole states in CI basis while keeping single-particle electron states to the single-particle ground state doublet, while in (b) when 48 single-particle electron states were included in CI basis while the number of single-particle hole states were kept to the single-particle ground state doublet.
 }
	\label{fig:Teo4X}
\end{figure}
\section*{Appendix III. Excitonic fine structure and effect of CI basis choice}
\label{appendixIII}

In Fig.~\ref{fig:Teo4X} we show the computed evolution of the four components (two bright and two dark for $\epsilon_{xx}\simeq 0$) of ground state exciton with $\epsilon_{xx}$ for two different CI basis choices. Similarly as observed in Ref.~\onlinecite{Yuan2018a}, one of the dark components (D1) crosses in energy both bright exciton components B0 and B1 and is vertically polarized. Since that changes the ordering of exciton states, one has to carefully monitor the results computed by CI in order to correctly attribute the computed bands to each of the four exciton states. We also note that the position of the almost degenerate dark states in absence of strain is about 60--70~$\mu$~eV below the bright states, which is close to, but somewhat lower than the experimentally measured values of about 110~$\mu$~eV for similar QDs.~\cite{Huber2019}

\begin{figure}[!ht]
	\centering
    \includegraphics[width=85mm]{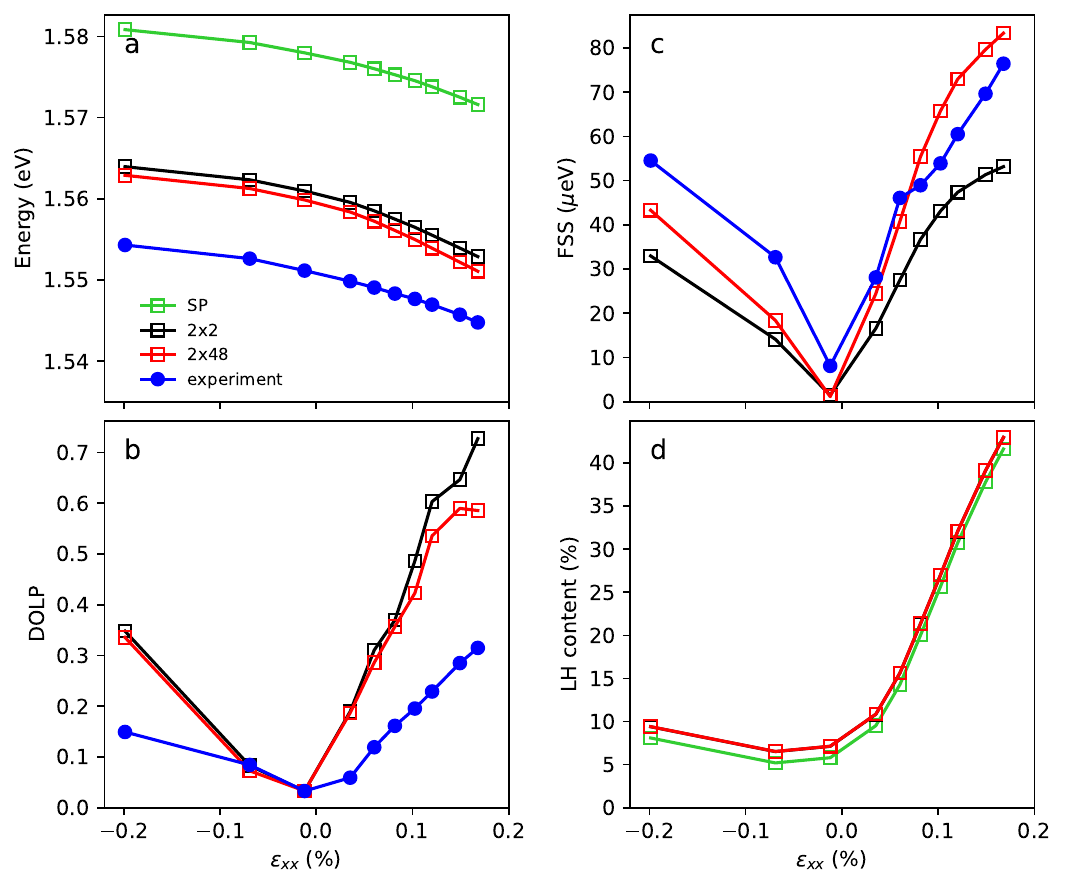}
	\caption{Same as in Fig.~\ref{fig:ExpVsCI} but for a different choice of CI basis states. In panels (a)--(c) we show the experiment (blue circles) and in (a)--(d) calculations performed without considering Coulomb interaction between electrons and holes (green squares) and those obtained using CI with single-particle bases of $2\times2$ (black squares) and $48\times2$ (red squares) electron~$\times$~hole single-particle states.}
	\label{fig:ExpVsCIel}
\end{figure}
In Fig.~\ref{fig:ExpVsCIel} we show the theory prediction of the effect of the applied stress on energy, DOLP, FSS, and LH content for the case in which we have varied the number of single-particle electron states in the CI basis while keeping the single-particle hole states fixed to the ground state doublet.
\begin{figure}[!b]
	\centering
    \includegraphics[width=85mm]{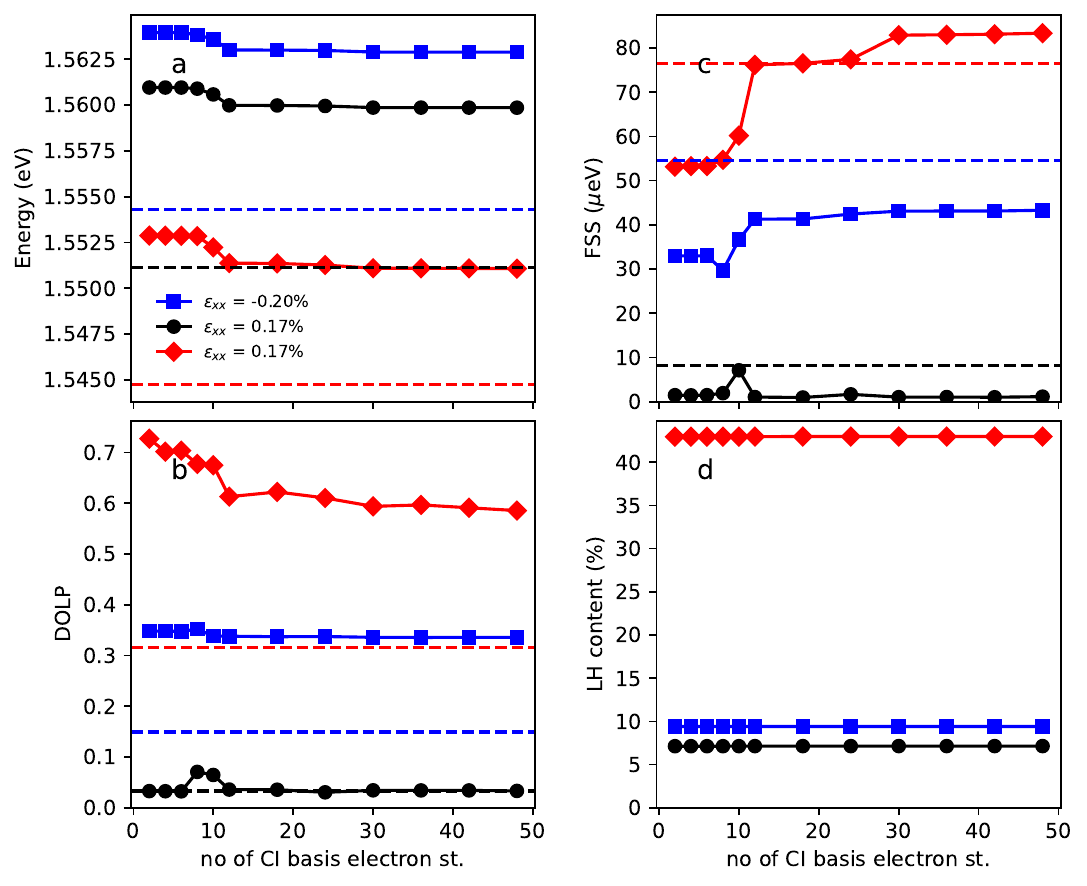}
	\caption{(a)--(d) Dependence of our ${\bf k}\cdot{\bf p}$+CI calculations results on the number of single-particle electron states included in the CI basis for three different configurations with the indicated values of $\epsilon_{xx}$. (Only the two ground-state single-particle hole states are included, see main text). The broken horizontal lines in (a)--(d) show the corresponding experimental values.}
	\label{fig:ExpVsCIconvel}
\end{figure}

Furthermore, in Fig.~\ref{fig:ExpVsCIconvel} we show the converge of our ${\bf k}\cdot{\bf p}$+CI calculations with increasing number of single-particle electron states in CI basis with single-particle hole states fixed to the ground state doublet. The convergence is shown for three distinct values of $\epsilon{xx}$ corresponding to Fig.~\ref{fig:ExpVsCIel}.

We see that the agreement, particularly for tensile $\epsilon_{xx}$, between theory and experiment for FSS is better when the number of single-particle electron states in the CI basis is varied while keeping the single-particle hole states fixed to the ground state doublet. This points to the relative importance of excited electron single-particle states for FSS. However, a poorer agreement is seen for DOLP. That is due to importance of HH-LH mixing for DOLP. Naturally, HH-LH mixing is larger for single-particle hole states, than for electrons. On the other hand, the convergence of CI for varying the number of single-particle electron states in the CI basis keeping the single-particle hole states fixed to the ground state doublet is considerably improved in Fig.~\ref{fig:ExpVsCIconvel} compared to Fig.~\ref{fig:ExpVsCIconv}. That is because of the energy spacing between single-particle electron states, which is larger than for single-particle hole states. Due to that, the content of the higher energy excited single-particle electron states in exciton is smaller, leading to more rapid convergence of CI. 

\end{document}